\documentclass{pas}

\usepackage{multirow}
\usepackage{xcolor}
\usepackage{orcidlink}
\usepackage{gensymb}
\newcommand{\kms}{km\,s$^{-1}$}
\newcommand{\hi}{H{\sc i}}

\newcommand\psj{\ref@jnl{PSJ}}

\begin{document}

\lefttitle{Publications of the Astronomical Society of Australia}
\righttitle{L. Cortese et al.}

\jnlPage{1}{4}
\jnlDoiYr{2021}
\doival{10.1017/pasa.xxxx.xx}

\articletitt{Research Paper}

\title{MAUVE: Cold neutral gas in the outflow of NGC\,4383 and evidence for a fountain flow}

\author{\sn{L.} \gn{Cortese}$^{1}$, 
\sn{A.~B.} \gn{Watts}$^{1}$, 
\sn{J.} \gn{Sun}$^{2}$, 
\sn{S.} \gn{Sankar}$^{1}$, 
\sn{B.} \gn{Catinella}$^{1}$, 
\sn{T.} \gn{Brown}$^{3,4}$, 
\sn{A.} \gn{Boselli}$^{5}$,
\sn{P.} \gn{J\'achym}$^{6}$,
\sn{T.} \gn{Kolcu}$^{7}$,
\sn{S.} \gn{Thater}$^{8}$,
\sn{J.} \gn{van de Sande}$^{9}$,
\sn{V.} \gn{Villanueva}$^{10}$
}

\affil{$^1$International Centre for Radio Astronomy Research, University of Western Australia, M468, 35 Stirling Highway, Crawley, WA 6009, Australia, $^2$Department of Physics and Astronomy, University of Kentucky, 506 Library Drive, Lexington, KY 40506, USA, $^3$National Research Council of Canada, Herzberg  Astronomy and Astrophysics Research Centre, 5071 W. Saanich Rd. Victoria, BC, V9E
2E7, Canada, $^4$Department of Physics \& Astronomy, University of Victoria, Finnerty Road, Victoria, BC V8P 1A1, Canada, $^5$Aix Marseille Univ, CNRS, CNES, LAM, Marseille, France, $^6$Astronomical Institute of the Czech Academy of Sciences, Boční II 1401, 141 00, Prague, Czech Republic, $^7$School of Physics and Astronomy, University of Nottingham, University Park, Nottingham NG7 2RD, UK, $^8$Department of Astrophysics, University of Vienna, Türkenschanzstrasse 17, A-1180, Vienna, Austria, $^9$School of Physics, University of New South Wales, Sydney, NSW 2052, Australia, $^{10}$Instituto de Estudios Astrof\'isicos, Facultad de Ingenier\'ia y Ciencias, Universidad Diego Portales, Av. Ej\'ercito 441, Santiago 8370191, Chile}

\corresp{L. Cortese, Email: luca.cortese@uwa.edu.au}



\begin{abstract}
We present a multiphase study of the star-formation-driven outflow in the Virgo galaxy NGC\,4383, combining ALMA CO(2–1) data with deep MeerKAT \hi\ imaging and MUSE spectroscopy obtained as part of the Multiphase Astrophysics to Unveil the Virgo Environment (MAUVE) program. Our previous work revealed a spectacular ionised outflow, but the effect of the outflow on the cold phase remained unclear. Our analysis shows that potentially outflowing molecular gas is detected only within the inner $\sim$1 kpc above the disc, where CO clouds exhibit disturbed kinematics and spatial correspondence with the ionisation cone. At larger heights, the CO surface brightness rapidly drops, indicating that the molecular phase contributes little to the mass of outflowing gas. In contrast, the \hi\ distribution shows plumes a few kiloparsecs above the disc that are aligned with the ionised cone, and complex kinematics suggestive of parts of the atomic phase being entrained in the outflow. However, the extended and warped \hi\ disc associated with NGC\,4383 complicates the unambiguous identification of outflowing atomic gas and, most importantly, the quantification of outflowing mass and loading factor. Independent support for a cold component in the outflow comes from dust extinction features associated with the outflow and coincident with \hi\ plumes. Despite significant uncertainties in the estimate of the mass of cold gas associated with the outflow, these results suggest that the atomic phase likely dominates the cold outflow above $\sim$1 kpc. The observed cold gas velocities remain below the velocities of the ionised phase, suggesting that NGC\,4383 does not host a large-scale escaping wind but more likely a galactic fountain, in which feedback redistributes material within the halo and regulates ongoing and future star formation.
\end{abstract}


\maketitle

\section{Introduction}
How gas cycles in and out of galaxies is key for understanding their life cycle. Indeed, it is the continuous circulation of baryons in the halo that truly regulates the star formation history of galaxies \citep{tumlinson17,saintonge22}. 

Historically, this cycle was seen as composed of two relatively distinct processes: inflows were thought to be associated with gas accretion from satellites or cold filaments (e.g., \citealp{keres05,sancisi}), while outflows were linked to feedback from supernovae or accreting supermassive black holes (e.g., \citealp{hopkins12}). In this classical picture, inflows and outflows can be seen as parts of the baryon cycle driven and regulated by very different physical mechanisms.

However, growing evidence in the local Universe suggests that this dichotomy is overly simplistic, at least at current epochs. On the one hand, feedback-driven winds fail to reach the escape velocity of their host galaxies, especially when driven by star formation, making it unlikely that they permanently eject large masses of gas into the intergalactic medium (e.g., \citealp{leroy15,concas17,mcquinn19,robertsborsani19,marasco23,watts24,ciraulo25}). On the other hand, direct evidence for cold filamentary inflows has never been obtained in the local Universe, and the accretion of satellites does not seem to be able account for the sustained star formation observed in most galaxies at late times (e.g., \citealp{sancisi,diteodoro14}). Instead, a galactic fountain model, in which gas cycles between the disc and halo before returning to fuel future star formation, is gaining significant traction \citep{fraternali08,marinacci10,afruni23}. In this view, outflows and inflows are not separate, but different phases of the same circulation process.

Probing this cycle observationally requires tracing the different gas phases of the interstellar medium (ISM) as they leave and re-enter galactic discs. While the warm gas component is relatively straightforward to map thanks to recombination lines in the rest-frame optical spectrum, this phase is not expected to dominate the mass budget of outflows (e.g., \citealp{veilleux20} and references therein). Instead, most of the mass is expected to reside in colder phases, either molecular or atomic (e.g. \citealp{bolatto13,cicone14,leroy15,fluetsch19,avery22}). Molecular gas traced by CO has been observed in a number of nearby normal star-forming systems, but typically only within the inner kiloparsec \citep{bolatto13,leroy15,noon23}, consistent with expectations from numerical models (e.g., \citealp{Vijayan24}). Atomic hydrogen, by contrast, is expected to be the dominant component on kiloparsec scales, but direct evidence for its participation in galactic outflows is still very limited (e.g., \citealp{mcclure18,martini18,heyer25}). This is due to a combination of sensitivity and resolution challenges, as well as the intrinsic complexity of \hi\ morphology, which often includes thick discs and warps that complicate the identification of extraplanar material \citep{marasco19} and might systematically overestimate any quantification of the mass of gas in process of leaving (or re-entering) the star-forming disc.  

Therefore, progress in this field requires a substantial increase in the number of galaxies for which multiphase outflows can be studied at sufficient spatial resolution to disentangle disc and extraplanar gas. To this end, we are conducting the MAUVE project (Multiphase Astrophysics to Unveil the Virgo Environment\footnote{\url{mauve.icrar.org}}), which is designed to trace the baryon cycle of galaxies as they fall into the Virgo cluster. At the time of writing, MAUVE is collecting observations from the Very Large Telescope/Multi Unit Spectroscopic Explorer (VLT/MUSE), the Atacama Large Millimeter Array (ALMA), and the Hubble Space Telescope to characterise the gas cycle in unprecedented detail in 40 galaxies in the Virgo cluster, achieving spatial resolutions ranging from $\sim$100 to $\sim$6 pc scales, depending on the facility \citep{mauvemsg}. This approach will allow us to probe the baryon cycle and determine how it is affected  by the cluster environment at different stages of infall into the cluster. One of the primary objectives of MAUVE is to reveal how star formation is quenched in satellites after the outer gas disc has been stripped \citep{boselli06,cortese21}. MAUVE has already uncovered striking examples of warm ionised, outflows at different stages of cluster infall, ranging from NGC\,4383 \citep{watts24}, which has only recently begun its descent into Virgo, to the post-pericentre, \hi-deficient system NGC 4064 \citep{attwater25}. 

In this paper, we present the first MAUVE multiphase study of a star-formation-driven outflow, focusing on NGC\,4383. This galaxy is the most actively star-forming galaxy in the whole MAUVE sample, based on its offset from the star-forming main sequence. Following \cite{leroy19}, NGC\,4383 has an integrated stellar mass of $\sim$10$^{9.4}$ M$_{\odot}$ and a current star formation rate of $\sim$1 M$_{\odot}$ yr$^{-1}$, positioning it $\sim$0.5 dex above the main sequence. According to \cite{chung09}, the galaxy has an inclination of 60 degrees with respect to the line of sight. 

Our MAUVE-MUSE observations revealed a powerful star-formation-driven, ionised outflow extending several kiloparsecs above the disc of the galaxy \citep{watts24}, while earlier single-dish and interferometric studies established the presence of an extended and warped \hi\ disc \citep{chung09}. Here, we combine ALMA CO(2–1) observations with MeerKAT \hi\ imaging and MUSE spectroscopy to jointly investigate the molecular, atomic, and ionised gas components of the outflow. Our aim is to evaluate the role of the cold phases in the baryon cycle of NGC\,4383 and to test whether its outflow is better described as a classical galactic wind or as part of a fountain-like circulation.

This paper is organised as follows. Section~\ref{sec2} describes the observations and data reduction. Section~\ref{sec3} presents the analysis of the molecular and atomic gas properties of NGC~4383, with direct comparison to the ionised component. Section~\ref{sec4} discusses the implications for gas cycling and feedback in NGC\,4383, and Section~\ref{sec5} summarises our conclusions. Throughout this work we adopt a distance to NGC\,4383 of 16.5 Mpc \citep{mei07,cantiello24}, for which 1 arcsecond corresponds to 80 pc.

\section{Data}
\label{sec2}
\subsection{ALMA CO~\,(2--1) data}
The ALMA CO~\,(2--1) observations of NGC\,4383 were acquired as part of the MAUVE-ALMA program (2023.1.00026.S, P.I. J.~Sun), which builds on the Virgo Environment Traced in CO (VERTICO) survey \citep{brown21}.
These data incorporate observations with both the ALMA 12m array and the Atacama Compact Array (ACA) to recover emission on all spatial scales. The observation field of view covers the star-forming inner disc of the galaxy out to $3{-}4$~kpc along the major axis. The spectral tuning covers the full velocity range of the $^{12}\mathrm{CO}\,(2{-}1)$ line emission at a native velocity resolution of $0.8$~km s$^{-1}$.

The data reduction scheme closely follows \citet{leroy21} and will be described in full detail in J.~Sun et al. (in preparation). Briefly, we start with calibrated visibility data supplied by the observatory and verify that no obvious artefacts are present in them. We then extract spectral channels within ${\pm}250$~km s$^{-1}$ of the systemic velocity for the $^{12}\mathrm{CO}\,(2{-}1)$ line, subtract the continuum in visibility space, rebin to a $2.5$~km s$^{-1}$ channel width, combine the 12m and 7m interferometry with the total power data, and image them following the PHANGS--ALMA imaging scheme \citep{leroy21}. For the imaging step, we adopt a Briggs weighting with a robustness parameter of 1.0 to achieve an optimal balance between sensitivity and resolution. Finally, we correct the cleaned data cube for the primary beam response pattern, convolve it to a round beam of $1$ arcsec in FWHM (80 parsec at the distance of Virgo), and feather it with single-dish data to obtain the final data cube.

We further produce CO line integrated intensity (moment-0) and velocity centroid (moment-1) maps from the data cube, adopting the same set of recipes as \citet{leroy21}. For the analyses in this work, we specifically use the `broad' moment-0 maps and the moment-1 maps `with a prior'. These maps 
are optimised to include diffuse CO emission (see \citealp{leroy21} for more details), important for comparisons with the large scale H$\alpha$ emission in the outflow from NGC\,4383. 

These data allow us to reach an integrated brightness sensitivity of $\sim$0.9 K \kms. This value corresponds to a molecular gas surface density of $\sim$5 M$_{\odot}$ pc$^{-2}$ if we assume a CO(2-1)-to-CO(1-0) flux ratio of 0.8 \citep{brown21} and a CO(1-0)-to-H$_{2}$ conversion factor typical of the Milky Way disc (i.e., 4.35 M$_{\odot}$ pc$^{-2}$ K \kms, \citealp{bolatto13b}). The adopted conversion factor agrees with the average value predicted by \cite{sun25} within the inner 1 kpc of NGC\,4383, derived from the stellar and star formation rate surface brightness profiles. 

\begin{figure*}
\centering
\includegraphics[width=1\linewidth]{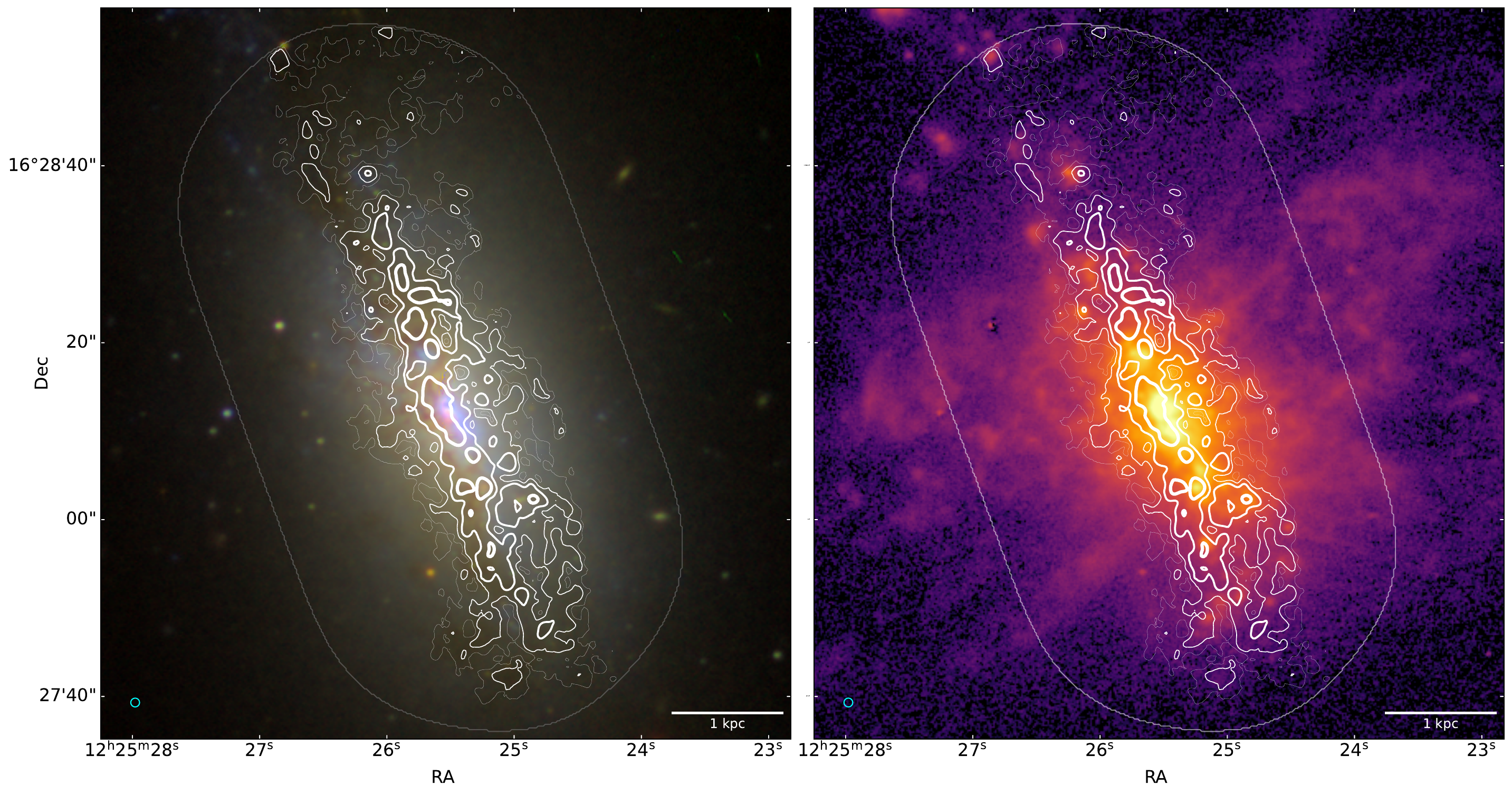}
\caption{{\bf The CO(2-1) morphology of NGC\,4383}. CO(2-1) intensity contours overlaid on a broad band $g$, $i$ and $z$ colour composite from NGVS (left) and H$\alpha$+[NII] net image from VESTIGE (right). Contours are shown at 0.9, 2.8, 9.2, 27.6 K \kms\ levels corresponding to $\sim$5, 15, 50, 150 M$_{\odot}$ pc$^{-2}$ assuming a Milky Way conversion factor (see Sec.~2.1). The footprint of the ALMA observations is shown in light grey, with the size of synthesised beam shown in the bottom left of each panel.}
\label{co_mom0}
\end{figure*}

\subsection{MeerKAT \hi\ data}
NGC\,4383 was observed with MeerKAT in April 2023  (ID: MKT-22034, P.I. Watts). The observation was taken
in L-band (856 MHz bandwidth centred at 1284 MHz) at 32k
spectral resolution (26.123 kHz wide channels, corresponding to 5.6 \kms\ at the frequency of NGC\,4383). 

Two tracks (rising and setting) were obtained on 03/04/23 and 17/04/23, alternating 30-min target scans with 2-min gain calibrator scans on J1120+1420. Bandpass calibrators J0408-6545 and J1939-6342 were observed at the start and end of each track, respectively. Observations were scheduled at least 1.5 hours after sunset to minimise solar interference, yielding a total observing time of 9h 12m (including overheads).

The two MeerKAT tracks were reduced using a custom version of the iDiA \textsc{processMeerKAT} pipeline\footnote{\url{https://github.com/spectram/pipelines/tree/HI-dev}} (see \citealp{maddox21}). Each track was calibrated and continuum-subtracted separately before being combined for \hi\ imaging. Standard flagging and cross-calibration were applied using the observed flux, bandpass and gain calibrators. 
Continuum subtraction was then performed using a combination of self-calibration and polynomial fitting of line-free channels. 
After subtraction, the two tracks were merged and imaged with \textsc{CASA} \citep{casa} using multiscale cleaning and robust=0 weighting. Following an initial shallow blind clean, a source mask was created with \textsc{SoFiA2} \citep{serra15,westmeier21} and applied for a deep clean to produce the final spectral cube.
The resulting cube has a synthesised beam of 12$\times$7.3 arcsec$^{2}$ ($\sim$960$\times$584 parsec$^{2}$), 26.123 kHz wide channels, a typical rms of 0.2 mJy beam$^{-1}$, $2$ arcsec pixel size and is primary beam corrected.

We extracted moment maps (integrated intensity, flux-weighted line-of-sight velocity and effective width) for the final data cube using the Python \textsc{maskmoment} package, developed by Tony Wong\footnote{\url{https://github.com/tonywong94/maskmoment}}. We required detections to exceed $2\sigma$ in at least two consecutive channels and to cover a minimum area of twice the synthesised beam. Since this study focuses on the high–surface-density regions of the \hi\ disc, we note that the precise mask definition for extracting moment maps does not affect our results. We favour the effective line width (defined as the ratio of the moment zero to the peak intensity multiplied by $\sqrt{2\pi}$) as it is more robust for weak signals \citep{heyer01,sun18}.

\begin{figure*}
\centering
\includegraphics[width=.99\linewidth]{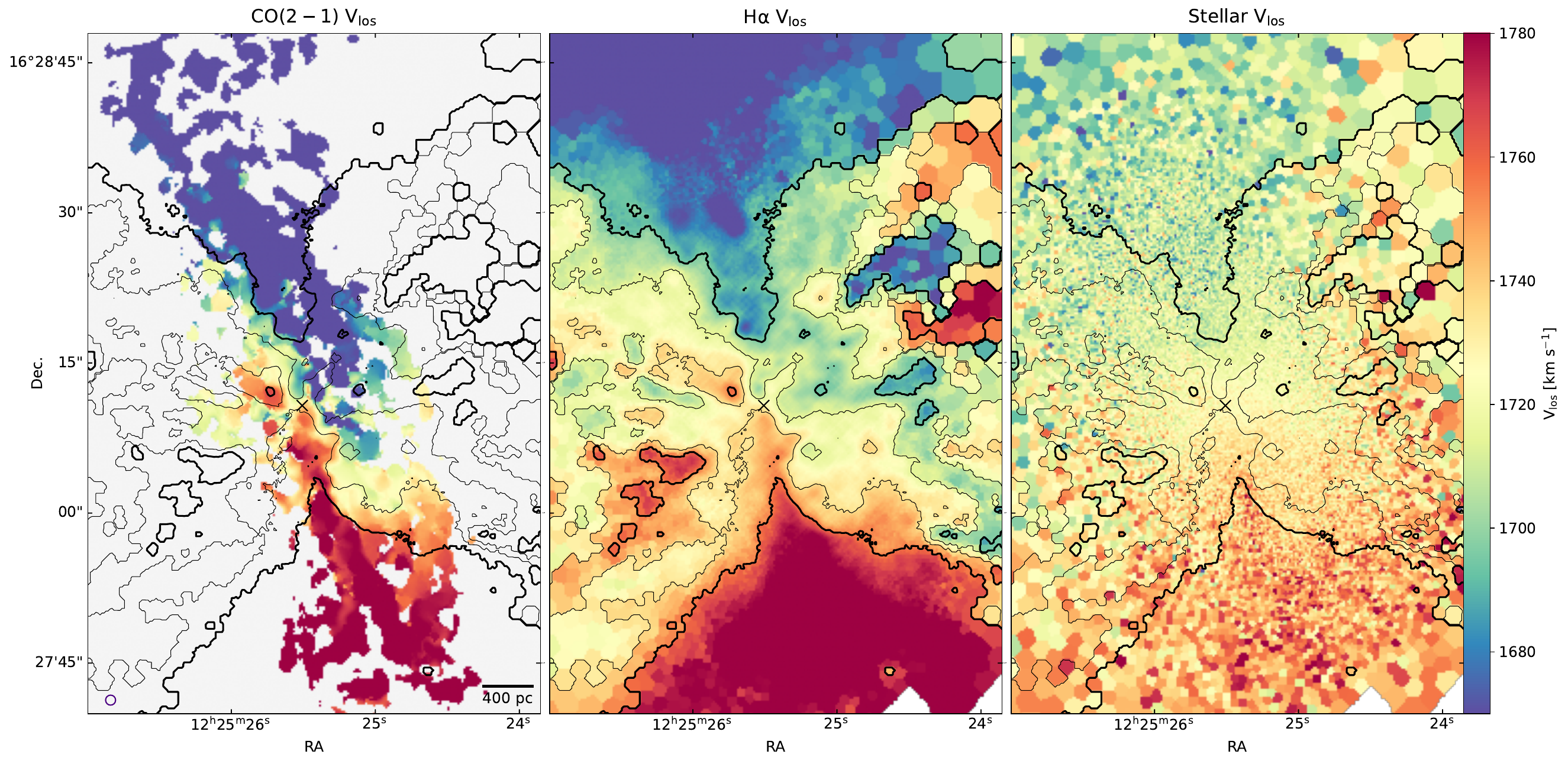}
\caption{{\bf The line-of-sight velocity fields of NGC\,4383}. Comparison between the CO(2-1) (left), the MUSE H$\alpha$ (middle) and the MUSE stellar (right) line-of-sight velocity fields. The black countours in both maps indicate H$\alpha$ velocities of 1695, 1755 (thick) and 1715 and 1735 (thin) \kms, respectively. The size of the ALMA synthesized beam is shown in the bottom left of the left panel. Fore reference, the ‘x’ marks the assumed kinematic centre of the CO emission.} \label{co_kin}
\end{figure*}

\subsection{MUSE and optical imaging data}
The MAUVE-MUSE observations of NGC\,4383 and their data reduction are presented in \citet{watts24}. For this work, the science-ready cube has been re-reduced using an updated version of \textsc{pymusepipe} (v2.28.2; \citealp{emsellem22}) as part of the MAUVE-MUSE internal data release v2 (see also \citealp{brown25}), but the differences between the two cubes are negligible (i.e. well within 1-2\% in flux). 

The emission-line flux and velocity maps used here were obtained with the new Galaxy IFU Spectroscopy Tool pipeline (nGIST; \citealp{ngist_paper,ngist_ascl}). nGIST builds on the GIST software developed by \citet{bittner19} and provides a flexible interface to the penalised pixel-fitting code \textsc{ppxf} \citep{ppxf1,ppxf2,ppxf3}.
The emission-line maps were generated by running nGIST on data Voronoi-binned to a target median S/N=20 in the 4800–7000 \AA\ wavelength range. This setting provides a good compromise between retaining the native resolution (seeing of $\sim$1.3 arcsec, corresponding to 100 pc at 16.5 Mpc) in the inner disc of the galaxy, where the high resolution is key for a direct comparison with the ALMA data, while ensuring robust continuum subtraction in the extraplanar regions. The extra-planar region is primarily used for the comparison with the \hi\ and this binning scheme guarantees that even the largest bin at the edge of the MUSE field of view is always smaller than the synthesised beam of the MeerKAT observations.   

Following the approach described by \citet{emsellem22}, we adopted the Medium-resolution INT Library of Empirical Spectra (MILES; \citealp{miles}) stellar population synthesis models (convolved to the same spectral resolution of the MUSE data) to first fit the stellar kinematics while masking emission lines, and subsequently fit the emission lines while keeping the stellar kinematic parameters fixed to the values obtained in the first step. We tie the line-of-sight velocity and velocity dispersion of emission lines in three groups: hydrogen lines (H$\alpha$, H$\beta$), low ionization lines (e.g.  [NII]$\lambda$6583, [SII]$\lambda\lambda$6716, 6731), 
and high ionization lines (e.g. [OIII]$\lambda$5007). All three groups exhibit consistent kinematics within the main body of the galaxy, while showing variations of up to a few tens of km s$^{-1}$ at larger distances from the disc, highlighting the complex kinematics of the ionised outflow. In the rest of this paper, we adopt the H$\alpha$ velocity field as our primary tracer of the ionised gas kinematics. However, we note that our conclusions remain unchanged if any of the other groups were used.

We complement our MUSE dataset with broad-band and narrow-band imaging covering the full region mapped by our MeerKAT data. The narrow-band H$\alpha\rm +[NII]$ net image was obtained as part of the Virgo Environmental Survey Tracing Ionised Gas Emission (VESTIGE; \citealp{vestige18,vestige23}), while $u$, $g$, $i$, and $z$ broad-band images were taken from the Next Generation Virgo Cluster Survey (NGVS; \citealp{ngvs}).

All velocity fields presented in this work are given in the Kinematic Local Standard of Rest (LSRK) reference frame and adopt the optical velocity definition.

\begin{figure*}
\centering
\includegraphics[width=.99\linewidth]{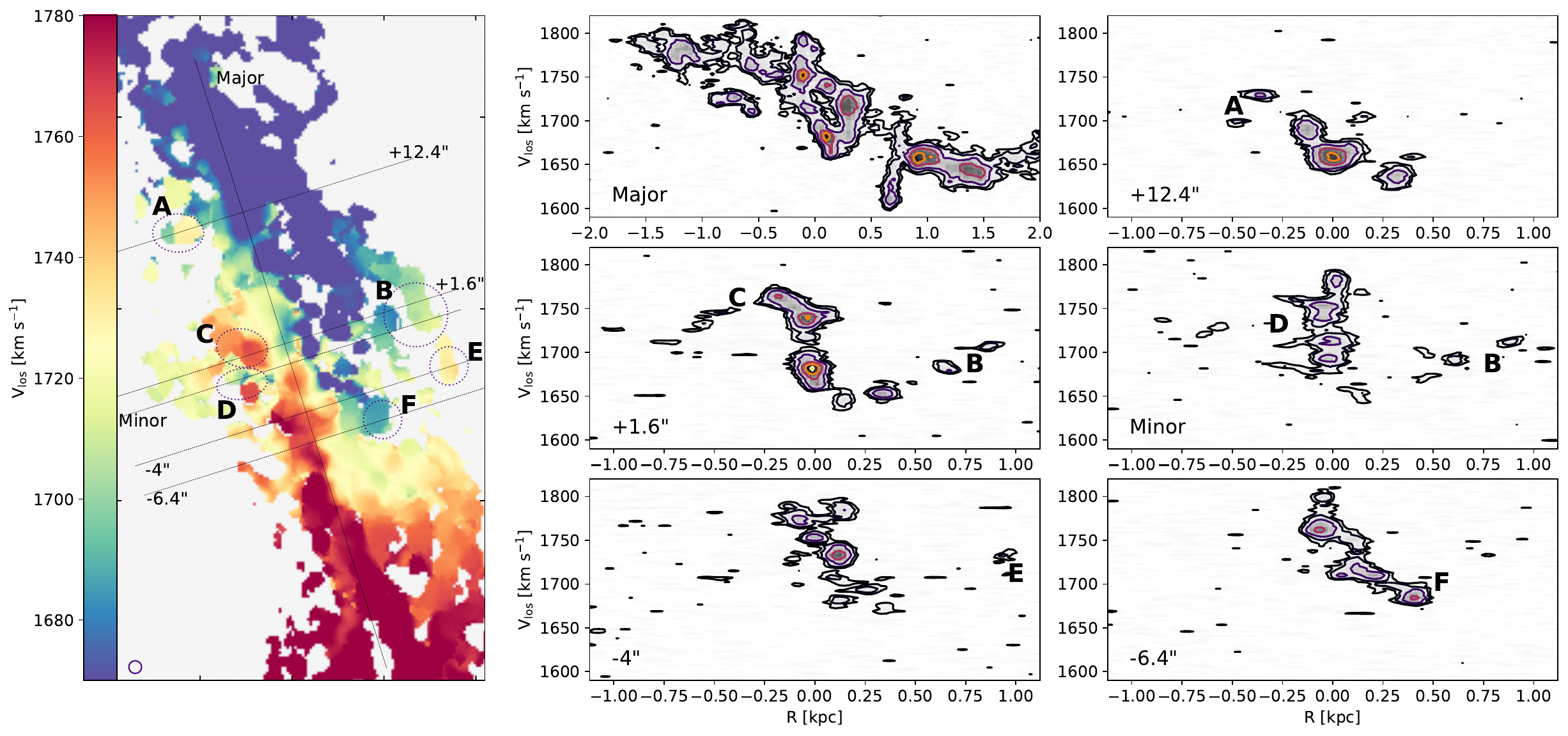}
\caption{{\bf Position-velocity diagrams for the CO(2-1) emission.} {\it Left:} CO(2-1) line-of-sight velocity field. The slits used to extract the PV diagrams are  overlaid. Interesting kinematic features are indicated from A to E. The size of the ALMA synthesized beam is shown in the bottom left. {\it Middle and Right:} PV diagrams. The position of each slit is indicated in the bottom-left corner of each sub-plot, with regions identified in the line-of-sight velocity field highlighted with the corresponding letter.}
\label{co_pv}
\end{figure*}

\section{Results}
\label{sec3}
\subsection{The impact of the starburst on the molecular gas disc}
We begin our analysis by examining the CO(2-1) properties of NGC\,4383. Figure \ref{co_mom0} shows the CO (2-1) integrated intensity contours overlaid on the NGVS optical broad-band colour composite (left) and on the VESTIGE H$\alpha$+[NII] (right) narrow-band images. The CO emission closely follows the distribution of the dust lanes visible in optical along the major axis of the stellar disc. When compared to the morphology of the ionised gas outflow, the two phases display strikingly different morphologies, with the CO emission extending no more than a projected distance of $\sim$ 1 kpc away from the plane of the galaxy. 
No CO emission is observed in the regions classified as `outflow-dominated' by \cite{watts24}, based on the velocity dispersion of the ionised gas in NGC\,4383.

Despite the absence of extraplanar CO emission extending several kiloparsecs above the disc,  and the lack of detailed correspondence between the 2D morphology of the CO and H$\alpha$+[NII] distributions, the high resolution of our CO data reveals the presence of small plumes and potentially detached clouds extending perpendicular to the disc out to distances of 0.5-1 kpc. To better assess the nature of these features, and to determine whether the molecular gas phase is at all affected by the outflow, the next step is to compare the kinematics of the ionised and molecular gas phases. 

In Figure \ref{co_kin} we compare the CO(2-1) (left), the MUSE H$\alpha$ (middle) and stellar (right) line-of-sight velocity maps. The same colourmap and velocity interval are adopted to facilitate a direct comparison between the three. To further guide the eye, we overlay the velocity contours of the H$\alpha$-emitting gas on top of the CO and stellar velocity maps, with the thick contours approximately marking where the velocity field of the ionised gas begins to exhibit the characteristic pattern of a rotating structure. It is clear that the stellar component shows significantly less rotation compared to both the molecular and ionised components. This is in line with the early-type optical morphology of this galaxy, which in the past was classified as an early-type spiral \citep{rc3,cortese12} or a lenticular galaxy \citep{trentham2002}.

Moving on to the comparison between the H$\alpha$ and CO maps, it is clear that they show both similarities and differences. Along the major axis, the velocity gradient of the CO appears steeper than the ionised gas. This is particularly evident on the northern side of the galaxy, where the CO disc reaches velocities lower than 1670 \kms\ ($\Delta V\sim$ 50 \kms) within $\sim$600\,pc from the centre, while the ionised phase reaches similar values only $\gtrsim$2 kpc from the centre. While this is at least partially due to asymmetric drift, it is also likely that the kinematics of the H$\alpha$ emission is more strongly impacted by the central starburst. The velocity field further indicates that the CO disc may be warped, resembling the \hi\ distortion reported by \citet{chung09}, especially along the south-west and north-east edges of the ALMA field of view.

\begin{figure}
\centering
\includegraphics[width=1.02\linewidth]{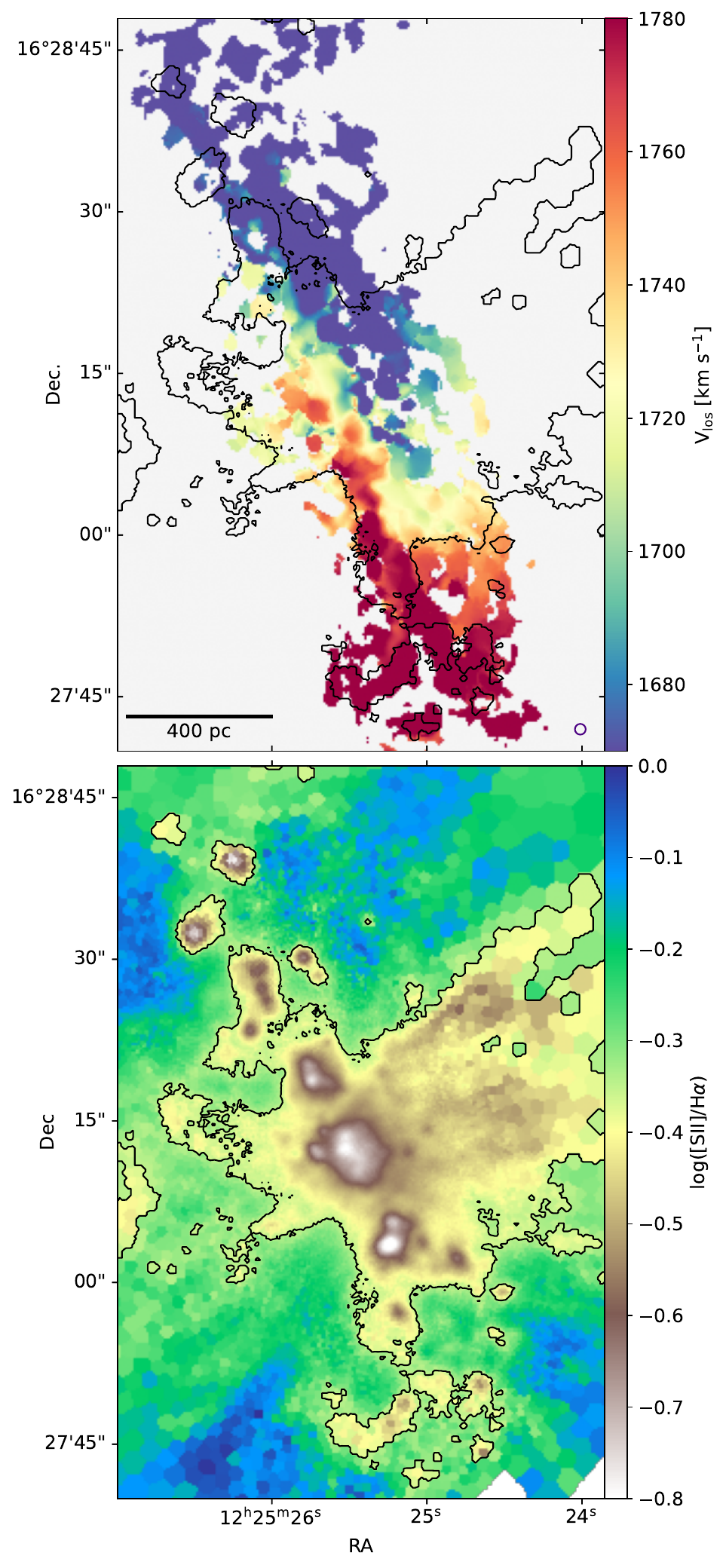}
\caption{{\bf [SII]/H$\alpha$ line ratio and CO(2-1) kinematics.} {\it Top:} CO(2-1) line of sight velocity field.  {\it Bottom:} [SII]/H$\alpha$ line ratio map. In both panels the black contour indicates $\log$([SII]/H$\alpha$)=$-$0.35}.
\label{co_sii}
\end{figure}

Focusing on the inner $\sim$1 kpc around the core of the starburst, the CO emission shows kinematics that deviates from regular rotation, most of which appear to qualitatively match features observed in the MUSE data. To better characterise the properties of these features, Figure~\ref{co_pv} presents a series of position-velocity (PV) diagrams extracted along the major axis and five different slits parallel to the minor axis to include the most prominent features in the CO velocity field. The width of each slit is 1 arcsecond. Given the complexity of the CO velocity field, we adopt the stellar position angle of 17.5\degree determined by \cite{watts24} as the kinematic position angle. We tested various ranges within $\sim$10\degree\ from this value and the results are not affected. Similarly we tested variations up to $\sim$2 arcseconds in the adopted kinematic centre and our results are unaffected.

Along the major axis, multiple velocity components are already evident along the line of sight, primarily within the inner 1 kpc. The presence of CO clouds kinematically detached from the disc becomes even more apparent when examining the slices taken parallel to the minor axis. Specifically, all features marked A–F exhibit velocities that deviate from the expectations for a rotating disc, suggesting that we may be observing CO clouds entrained in the outflow and displaced from the disc. Notably, features B, C, and F appear to correspond to similar structures in the ionised-gas velocity field. These deviations, however, remain within $\sim$100 \kms, substantially lower than the velocity range observed in the ionised phase at larger distances \citep{watts24}. 

\begin{figure*}[!t]
\centering
\includegraphics[width=1.0\linewidth]{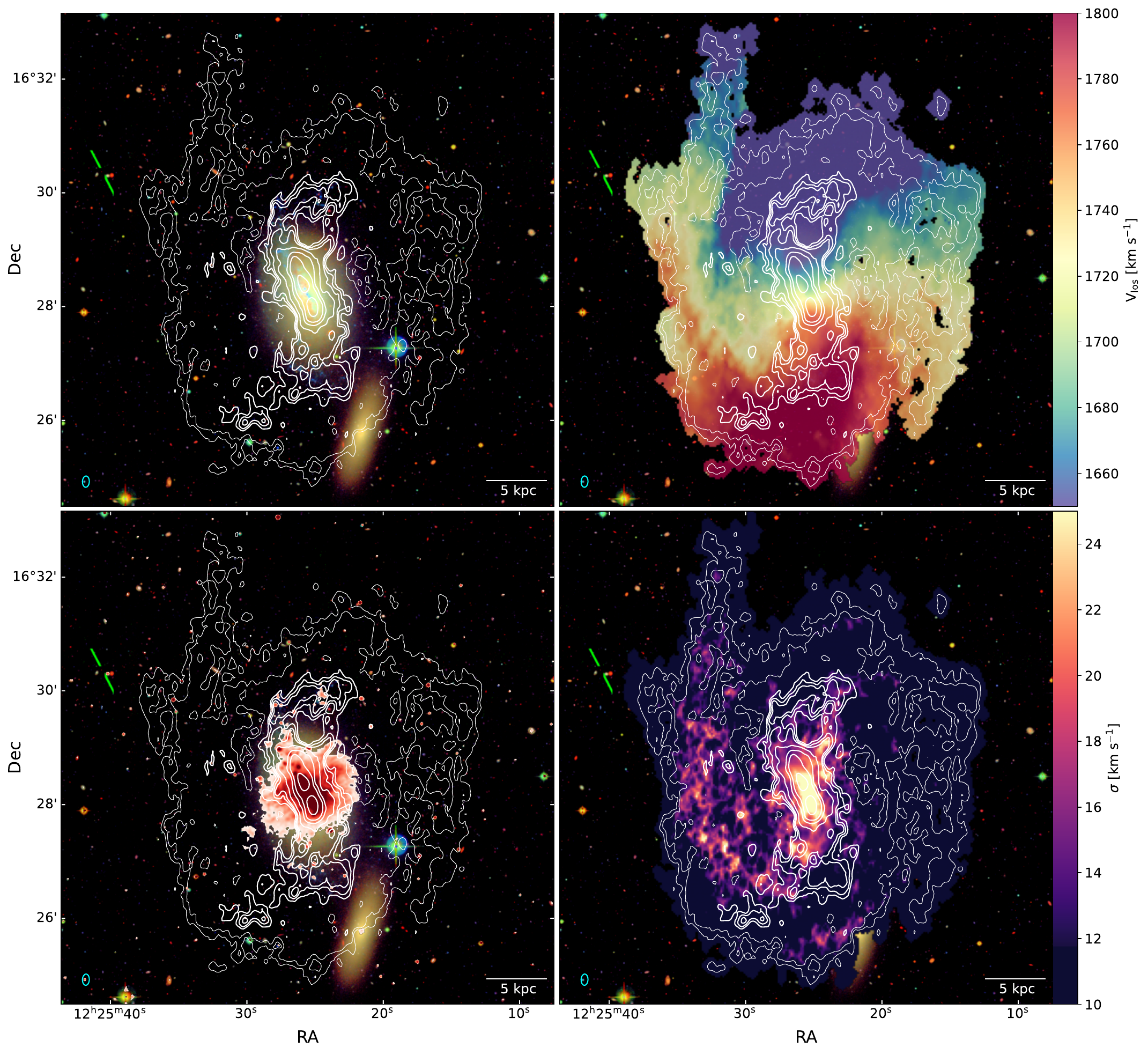}
\caption{{\bf The \hi\ disc of NGC\,4383 as revealed by MeerKAT}. \hi\ surface density contours (top-left) superposed on the \hi\ moment 1 (top-right), VESTIGE H$\alpha+[NII]$ narrow-band image (bottom-left) and \hi\ effective width (bottom-right) maps. Contours are shown at 1, 2 (thin), 5, 6, 7, 10, 15, 20 and 30 (thick) M$_{\odot}$ pc$^{-2}$. The background image shows the optical broad-band $u$,$g$,$i$ colour composite from NGVS. The synthesized beam of the \hi\ observations is shown in the bottom-left corner of each panel.}
\label{HIzoomout}
\end{figure*}

Intriguingly, the region where the CO kinematics shows the strongest evidence for potential outflow-related anomalies matches closely with the ionisation structure of the ISM as traced by the observed ratio of the [SII]$\lambda\lambda$6716, 6731/H$\alpha$ lines (similar results are obtained using other line ratios, e.g. [NII]/H$\alpha$). Specifically, a threshold of $\log$([SII]/H$\alpha$)$\sim$ -0.35 (solid contour both panels of Figure~\ref{co_sii}) encompasses very nicely the regions where the CO shows the strongest evidence for kinematics perturbations. 

However, it is unclear whether this correspondence implies a causal link between disturbed CO kinematics and the ionisation field, as both features are preferentially located close to the disc, where photoionisation from young stars is expected to dominate the ionisation budget. It is therefore likely that this correspondence partly reflects the fact that the disturbances in CO kinematics arise in regions where the ionised gas is still strongly influenced by radiation from the star-forming disc, while shock excitation becomes increasingly important at larger heights, as shown by previous studies of ionised winds (e.g. \citealp{sharp10}). At the same time, this correspondence may also represent indirect evidence for CO suppression (e.g. \citealp{Shimizu19}) or even complete destruction (e.g. \citealp{villanueva25}) driven by the intense radiation field of the starburst.


The evidence presented so far suggests that the CO disc is kinematically perturbed by the starburst occurring in the core of NGC\,4383, with clear indications that some of the CO-emitting gas is being displaced up to $\sim$1 kpc from the disc. While it remains challenging to confidently distinguish which features are associated with the outflow and which ones remain part of the disc, our findings already suggest that most of the CO is unlikely to travel far from the disc, or it is fully dissociated. 

The next step is to investigate whether a neutral wind exists at larger distances from the disc in the atomic phase. The substantially lower resolution of the \hi\ data makes this task even more challenging. This is apparent in Appendix~1, where we reproduce Figure~\ref{co_pv} at a resolution of 500 pc, closer to that of the \hi\ observations. It is clear that most, if not all, evidence of disturbed CO kinematics disappears or becomes significantly less pronounced.

\subsection{Searching for the atomic phase of the outflow}
\label{secHI}
The \hi\ properties of NGC\,4383 are considerably more complex than those of the molecular phase. \citet{chung09} reported the presence of an unusually extended \hi\ disc when compared to the size of the optical disc (i.e., over 4 times larger than the optical one, measured at 25 mag arcsec$^{-2}$ in B-band), suggesting a possible accretion event as its origin. It is therefore useful to begin by zooming out and revisiting the large-scale \hi\ distribution and kinematics of this galaxy before focusing on the region dominated by the ionised outflow. In Figure~\ref{HIzoomout}, we compare the \hi\ properties revealed by our MeerKAT data with the optical morphology (top left) and with the H$\alpha$ emission (bottom left). The right panels show the \hi\ line-of-sight velocity ($V_{los}$, top) and effective velocity width ($\sigma$, bottom).   

\begin{figure*}[!t]
\centering
\includegraphics[width=1\linewidth]{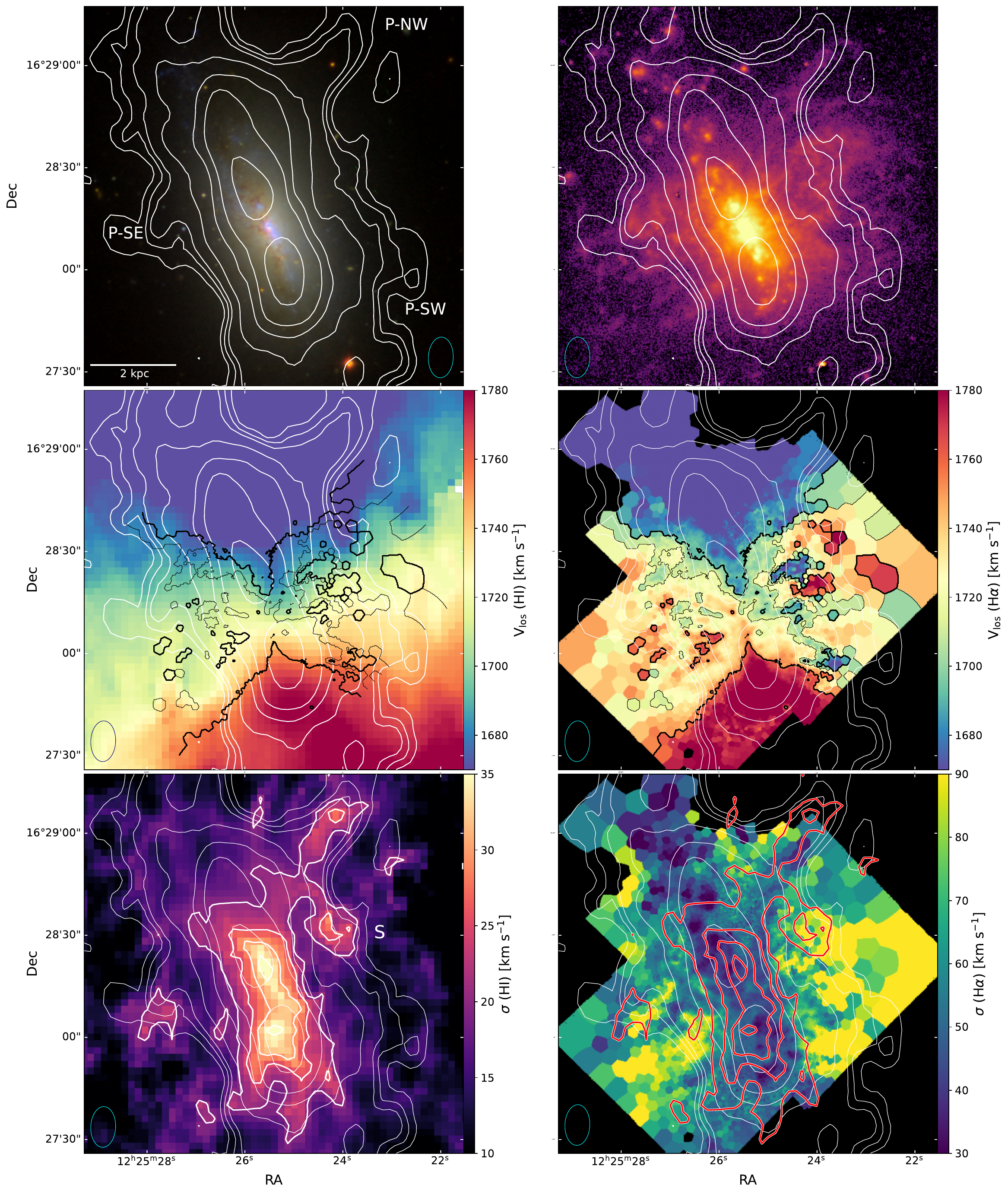}
\caption{{\bf The \hi\ and ionised gas outflow of NGC\,4383}. The \hi\ surface density contours superposed on the NGVS broad-band colour image (top-left), VESTIGE H$\alpha$+[NII] narrow-band image (top-right), \hi\ moment 1 map (middle-left), MUSE H$\alpha$ line-of-sight velocity map (middle-right), \hi\ effective line width (bottom-left) and H$\alpha$ velocity dispersion (bottom-right). The black contours in the middle row indicate H$\alpha$ line-of-sight velocities of 1695, 1755 (thick) and 1715 (thin) \kms. In the bottom row the white (left panel) and red (right panel) contours show \hi\ effective width at 20, 25, 30, 35 \kms\ level, respectively.}
\label{HIzoomin}
\end{figure*}

\begin{figure*}
\centering
\includegraphics[width=.99\linewidth]{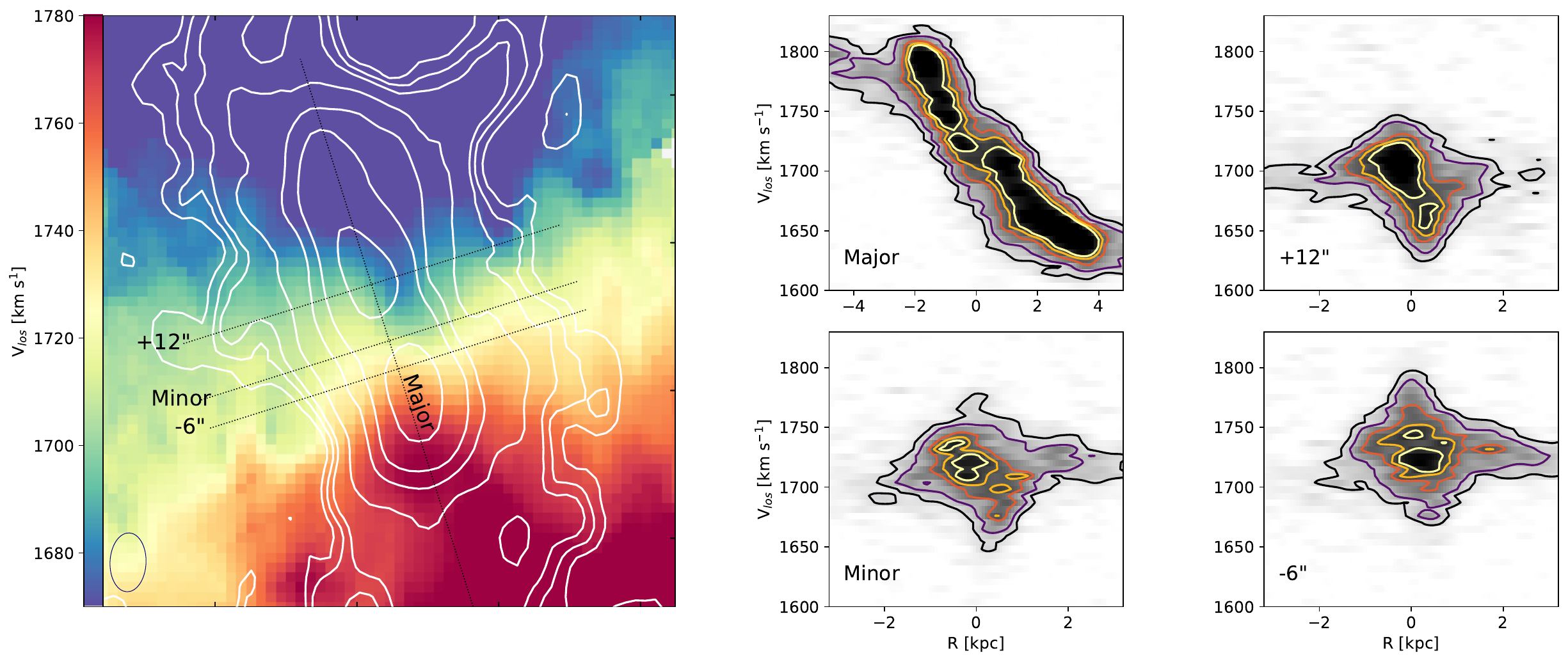}
\caption{{\bf Position-velocity diagrams for the \hi\ emission.} {\it Left:} \hi\ line of sight velocity field, with HI surface density distribution shown as the white contours (as in Figure~\ref{HIzoomin}). The black lines mark the slits used to extract the PV diagrams: along the major and minor axes, and offsets of 12 arcsec north and 6 arcsec south of the major axis. {\it Middle and Right}: The PV diagrams extracted along the corresponding slits shown in left panel.}
\label{HIpv}
\end{figure*}

The spatial distribution and kinematic properties of the \hi\ reveal the complexity of the gas reservoir of NGC\,4383. While the overall velocity field exhibits clear signs of rotation, the outer regions of the disc deviate from regular rotational motion. The most prominent feature is located at the east/north-east edge of the \hi\ distribution, where anomalous kinematics and an increase in $\sigma$ are observed. These characteristics likely result from the presence of multiple velocity components in the \hi\ and/or an irregular warp.

All of these features were previously noted by \citet{chung09}, and our data provide a higher-resolution view. However, the focus of this paper is on the inner, outflow-dominated regions, which were not accessible in earlier \hi\ interferometric studies. Within the inner $\sim$5 kpc, where the \hi\ surface density exceeds $\sim$5 M$_{\odot}$ pc$^{-2}$, the two-dimensional distribution of \hi\ resembles the typical structure associated with $m=2$ spiral patterns. Galaxy Evolution Explorer (GALEX) ultraviolet imaging \citep{watts24}, as well as H$\alpha$ (Figure~\ref{HIzoomout}, bottom-left panel), indicate recent star formation associated with these two `arms'. The \hi\ surface density increases significantly toward the centre, reaching values above 30 M$_{\odot}$ pc$^{-2}$ in the main stellar body. This suggests that the \hi\ disc is highly inclined, and then warps to a less inclined configuration at larger radii. This complex kinematic structure, combined with the abundance of \hi\ across the entire region of the ionised-gas outflow, makes it challenging to clearly identify any atomic hydrogen component unequivocally associated with the H$\alpha$ (and CO) outflow. In the remainder of the analysis, we thus focus on the inner $\sim$ 5 kpc of the \hi\ distribution, overlapping with the bulk of the ionised-gas outflow.

Figure~\ref{HIzoomin} presents a detailed comparison of the 2D intensity distribution (top), line-of-sight velocity (middle), and velocity dispersion (bottom) of the neutral atomic and ionised phases in NGC\,4383. The \hi\ exhibits two symmetric surface-density peaks along the major axis located roughly $\sim$1 kpc from the centre of the outflow, with a decrease in the inner $\sim$800 pc, where the brightest H$\alpha$ and CO emissions are observed. While this feature might be interpreted as evidence for central depletion by the outflow, the CO morphology and the fact that the ISM mass is dominated by molecular hydrogen suggest that the dip primarily reflects the predominance of the molecular phase in the central regions.

More intriguing is the morphology of the \hi\ along the minor axis, which appears to form plumes that sometimes follow the ionised-gas distribution. This is particularly notable at the northern edges of the H$\alpha$ outflow: to the east, the two phases align almost completely (labelled P-SE in top-left panel of Figure~\ref{HIzoomin}), whereas to the west, the \hi\ appears offset by $\sim$10 arcsec to the north relative to the ionised component (labelled P-NW). Moving to the southern envelope of the outflow, we see also a potential \hi\ plume extending to the west (labelled P-SW). While this might suggest the presence of atomic hydrogen associated with the extraplanar ionised-gas outflow, Figure~\ref{HIzoomin} represents only the central part of a more extended and complex \hi\ distribution, so caution is warranted in attributing the bulk of the observed emission to outflowing gas. What is already evident, however, is that the cold neutral ISM becomes predominantly atomic at distances of $\sim$1 kpc above the main disc. 

As with the CO, the next step is to examine in detail the kinematic properties of the \hi\ and compare them with those of the ionised phase of the outflow. In contrast to the CO, the \hi\ line-of-sight velocity field does not exhibit the same level of kinematic disturbance observed in H$\alpha$ (Figure~\ref{HIzoomin}, middle left), suggesting that either the \hi\ kinematics are only weakly affected by the outflow, or, more likely (due to the diffuse and extended distribution of the \hi), that any outflow-related effects on the moment 1 map are detectable only at spatial resolutions higher than those provided by our MeerKAT data.

The difference between the ionised and \hi\ kinematics becomes even more evident when comparing the line broadening of the two phases (bottom-row of Figure~\ref{HIzoomin}). While the H$\alpha$ velocity dispersion (corrected for instrumental broadening) increases markedly from the disc (below the MUSE spectral resolution of $\sim$50 \kms) to the outflow (above 90 \kms), the \hi\ shows the opposite trend: significant broadening (up to $\sim$35 \kms) is observed within the inner $\sim$1.5 kpc from the disc plane, but rapidly decreases to $\sim$15 \kms, characteristic of a cold rotating disc. The regions P–NW and P–SE coincide with areas where the ionised outflow shows intermediate velocity dispersions ($\sim$60 \kms), with both \hi\ plumes roughly corresponding to two of the only three locations outside the main disc where the \hi\ line width exceeds $\sim$20 \kms. Specifically, P–NW stands out clearly in the \hi\ line-width map, whereas P–SE shows enhanced broadening only near its tip. In addition to these two plumes, we identify a third region of enhanced \hi\ line broadening, not previously apparent in either the surface density or velocity maps. Labeled $S$ in the bottom left panel of Figure~\ref{HIzoomin}, this feature is projected $\sim$2 kpc above the disc plane, coinciding with one of the most complex regions of ionised-gas kinematics, where line-of-sight velocities of the H$\alpha$ line span more than 100 \kms\ within a single \hi\ beam.

To complete the characterisation of the \hi\ kinematics of the inner regions of NGC\,4383, in Figure~\ref{HIpv} we show the position–velocity diagram for the \hi\ along some of the same line-of-sights used for the CO. Although the lower spatial resolution of the \hi\ data restricts us to three locations along the minor axis, we find similar deviations from ordered kinematics to what observed for the CO. At distances beyond 2 kpc, the \hi\ shows velocities broadly consistent with disc rotation, whereas within the inner 2 kpc clear departures are observed: the southern part of the disc is redshifted and the northern part blue-shifted, consistent with the behaviour of both the H$\alpha$- and CO-emitting gas. Positive velocity residuals in the SE and negative in the NW side of the outflow are also recovered if we attempt to fit the whole \hi\ kinematics with a tilted-ring model using softwares such as 3D Barolo \citep{3dbarolo}. However, the exact amount of these residuals significantly depend on the assumptions made on the geometry of the disc and, given the complex 3D structure of the HI even in the outer parts of the gas disc, we prefer to show the data themselves without introducing any model-dependent assumption. 

In summary, while the limited spatial resolution of the \hi\ data hampers our ability to clearly separate atomic hydrogen associated with the outflow from the one belonging to the disc, we find clear evidence that part of the atomic hydrogen reservoir is directly affected by, and potentially entrained in, the ionised-gas outflow of NGC\,4383 within the inner $\sim$1-2 kpc from the plane of the disc and tantalising hints that it may extends up to at least 2.5 kpc, in particular at the edges of the cone of the ionised outflow.

\section{Discussion}
\label{sec4}
In this paper, we have presented new evidence that both molecular and atomic hydrogen in NGC\,4383 are directly affected by the outflow, with at least part of the cold gas reservoir likely entrained in the wind. While each phase contributes differently across spatial scales, our results highlight the multi-phase nature of the outflow.

\begin{figure}[t]
\centering
\includegraphics[width=0.95
\linewidth]{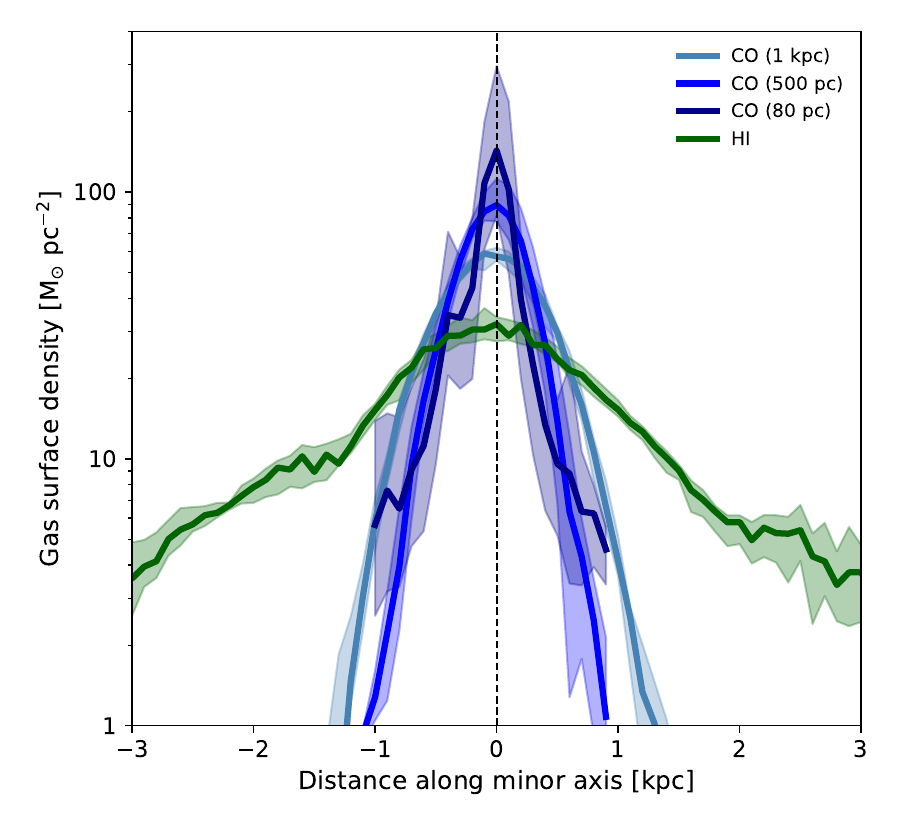}
\caption{{\bf Cold gas surface density profiles along the minor axis.}
\hi\ (green) and CO(2–1) (blue) surface density profiles extracted along the minor axis of NGC\,4383 using a slit 1.5 kpc wide. For the CO emission, different shades of blue indicate profiles derived from the original cube and from cubes imaged at 500 pc and 1kpc  spatial resolution to approximately match that of the \hi\ data.}
\label{profiles}
\end{figure}

The molecular component traced by the CO(2-1) transition is detected primarily within the inner kiloparsec from the heart of the starburst. Beyond this scale, the molecular hydrogen column density rapidly falls below our sensitivity limits, suggesting that the molecular phase contributes only marginally to the outflowing gas mass outside the central kiloparsec. This is clearly illustrated in Figure~\ref{profiles}, which compares the surface density profiles of the CO-emitting gas and the \hi\ extracted along the minor axis using a slit 1.5 kpc wide. For the CO emission, we additionally show profiles derived from cubes imaged to spatial resolutions of 500 pc and 1 kpc, chosen to approximately match the minor and major axes of the MeerKAT synthesized beam. Even at lower spatial resolutions, the CO emission does not extend significantly beyond $\sim$1 kpc.
Only if the $\alpha_{CO}$ conversion factor in the outflow would increase by $>$5 times the value assumed in this paper, the molecular phase could still contribute to a significant fraction of the outflow rate at larger distances. However, previous studies have generally found $\alpha_{CO}$ values for outflowing CO-emitting gas to be similar or lower than the Milky Way value (.e.g, \citealp{dasyra16,peraira24}).

The lack of a dominant molecular component at heights above a couple of kpc from the disc seems consistent with theoretical predictions from \citet{Vijayan24}, who suggest that in galaxies with average star formation rate surface densities comparable to or lower than the value observed for NGC\,4383 ($\log(\Sigma_{SFR})\sim-$0.27 M$_{\odot}$ yr$^{-1}$ kpc$^{-2}$, \citealp{watts24}), the bulk of the outflowing cold material should be in the atomic phase. Nevertheless, the possibility remains that individual dense molecular clouds survive and travel to larger distances, as those observed in the Milky Way \citep{diteodoro20} and Small Magellanic Cloud \citep{mcclure18},  but at levels below our current sensitivity and/or resolution. Deeper millimetre observations will be required to probe this regime.

Moving to the HI, we find tantalising evidence for a cold neutral component directly connected to the outflow. We identify extraplanar features that are both spatially and kinematically correlated with the ionised phase. Despite the improvement in spatial resolution compared to previous observations, our MeerKAT data still limit the extent to which these connections can be quantified, particularly given the large-scale complexity of the \hi\ distribution in NGC\,4383.

\begin{figure}[t]
\centering
\includegraphics[width=1.01
\linewidth]{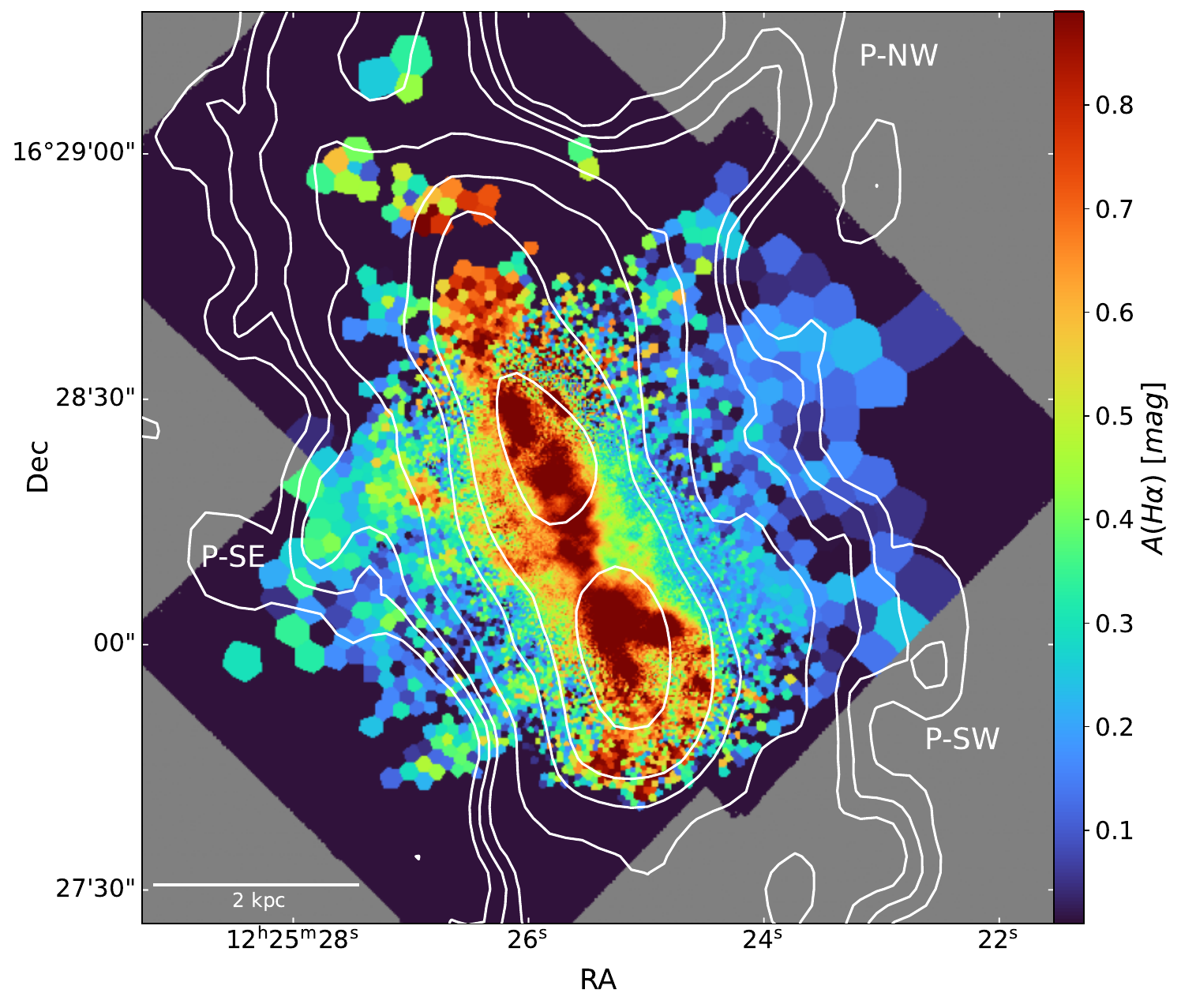}
\caption{{\bf Evidence of dust in the outflowing gas}. Map of H$\alpha$ attenuation for NGC\,4383. $A(H\alpha)$ is estimated from the Balmer decrement. It is clear that the line-emitting gas associated with the outflow is mixed with dust. White contours are from the \hi\ intensity map as in Figure~\ref{HIzoomin}.}
\label{Aha}
\end{figure}

An independent way to confirm the presence of cold material in the outflow is to trace dust associated with the neutral phase. Using our MUSE data, we constructed a gas-attenuation map from the H$\alpha$/H$\beta$ ratio. We consider only spaxels where the H$\beta$ line has a flux greater than 3$\times$10$^{-19}$ erg cm$^{-2}$ s$^{-1}$ and the ratio between line flux and the error on the line fit is larger than 5. We estimate $A(H\alpha)$ assuming Case B recombination, a temperature of 10$^{4}$ K and an electron density of 100 cm$^{-2}$ \citep{osterbrock06} and a \cite{fitzpatrick99} extinction curve. The gas attenuation map is shown in Figure~\ref{Aha}, with overlaid the \hi\ intensity contours. We find significant extinction (e.g., $A(H\alpha)\sim$0.3-0.5 mag) up to projected distances of at least  2.5 kpc, with a particularly strong correspondence with the P–SE feature, and tentative associations with parts of P–NW, as well as with more diffuse structures to the south-west (P-SW). This spatial match between dust attenuation and \hi\ strongly supports the interpretation that part of the atomic hydrogen is indeed entrained in the outflow, since dust grains and neutral gas are tightly coupled in the cold ISM and tend to trace the same structures. However, the correspondence is not always perfect. While this could be due to the different spatial resolution of the MUSE and MeerKAT observations, it is at least partially due to the fact that the Balmer decrement only traces dust in front of the line-emitting gas and not the full dust column density along the line of sight. 

While the bulk of the \hi\ emission in P–SE is likely associated with outflowing material, other regions may instead be dominated by extraplanar gas in the thick disc, not directly linked to the ongoing outflow, or  possibly by gas raining back toward the plane, or accreted during a recent interaction. This highlights an important caveat for direct estimates of mass-loading factors: such measurements often neglect the coexistence of gas currently entrained in the current outflow and extraplanar material not associated with the outflow that is just projected along the same lines of sight.

\begin{figure}[t]
\centering
\includegraphics[width=1.01
\linewidth]{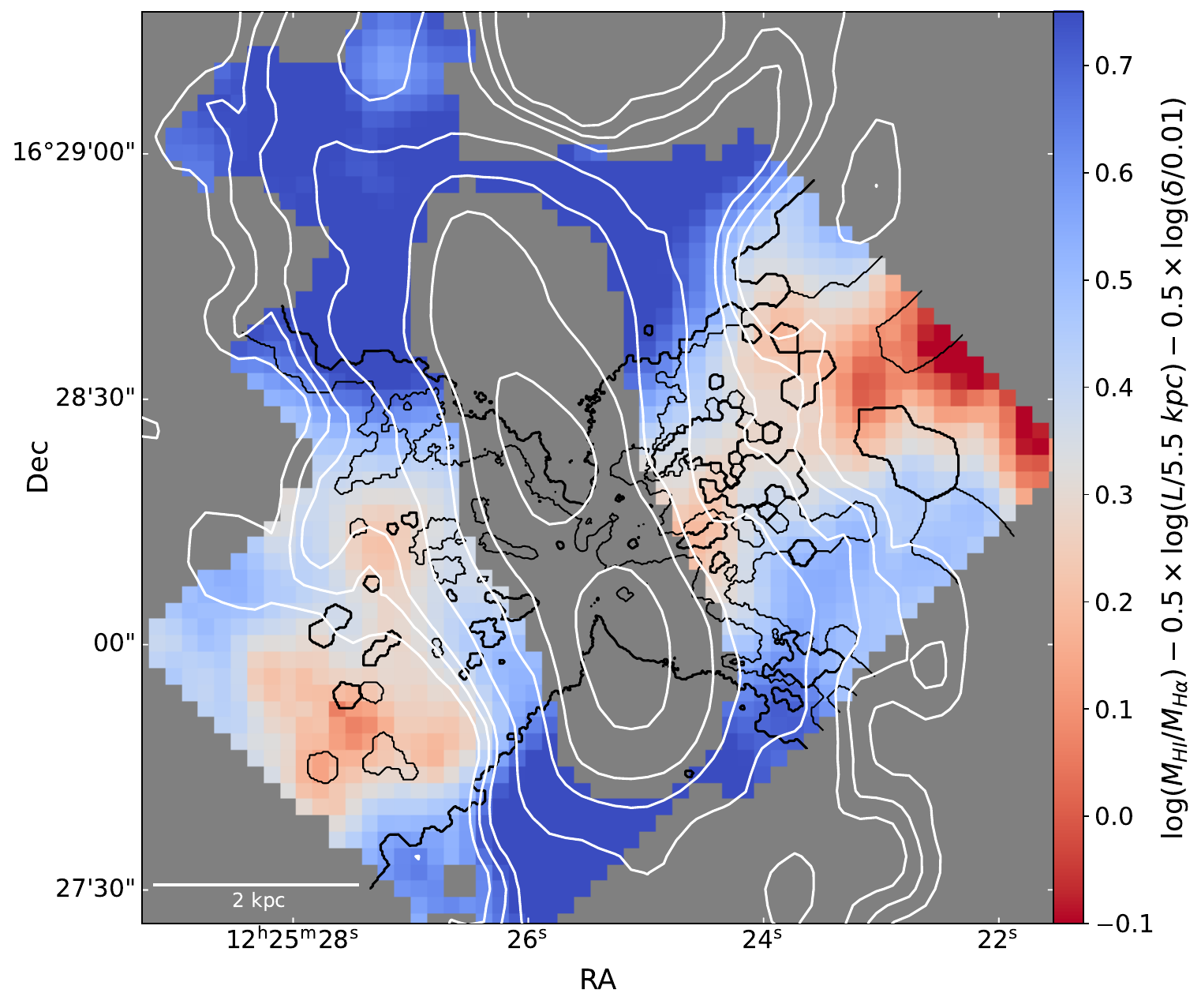}
\caption{{\bf The mass balance between H$\alpha$-emitting and neutral atomic gas}. Map showing the mass ratio of the \hi\ to the  H$\alpha$-emitting gas along the line of sight. The regions where the \hi\ surface density exceeds 8 M$_{\odot}$ pc$^{-2}$ are masked as these are clearly dominated by \hi\ in the disc. Black contours show H$\alpha$ iso-velocity regions as in Figure~\ref{co_kin} to identify outflow-dominated regions.
The estimate of the warm gas is highly uncertain and depends on both gas filling factor $\delta$ and the depth of the outflowing gas along the line of sight $L$, as indicated by the colour bar. White contours are the \hi\ intensity map as in Figure~\ref{HIzoomin}.}
\label{massratio}
\end{figure}

Indeed, it would be tempting to use these data to present an estimate of the mass outflow rate in the different phases of NGC\,4383. However, given the large uncertainty in any estimate of the amount of \hi\ associated with the outflow, as well as in the geometry of the outflow and filling factor of the ionised gas (critical parameters for the warm gas mass outflow rate), any claim on mass ouflow and loading factor could be easily off by an order of magnitude. Thus, here we prefer to use a more conservative approach and investigate what we can learn from these data despite these large uncertainties.

First, any \hi\ component associated with the outflow shows a projected velocity lower than the H$\alpha$: where the H$\alpha$ has line-of-sight velocities exceeding 100 \kms, the associated \hi\ rarely departs from systemic by more than $\sim$50 \kms. Of course, there could be projection effects, but it is unlikely that the geometry of the outflow in the two phases is enough to account for this. However, the limited spatial resolution of the \hi\ observations may prevent the identification of small-scale gas clumps with velocities closer to those of the ionised phase. Second, the comparison of gas surface densities alone already suggests that the cold component may dominate the mass budget. Indeed, the ionised outflow mass estimated by \citet{watts24} of M$({H\alpha})$=5.1 $\times$ 10$^{7}$ M$_{\odot}$  over an area of 24.4 kpc$^{2}$, implies an average ionised gas surface density of $\sim$2.1 M$_{\odot}$ pc$^{-2}$, well below the typical \hi\ surface density in the regions coincident with the outflow, which ranges between 5 and 10 M$_{\odot}$ pc$^{-2}$. This would imply that the neutral gas mass is at least comparable to, and likely exceeds, the mass in the ionised phase.

However, our results have highlighted how the two phases are not always strictly co-spatial, with the \hi\ preferentially located toward the edges of the ionised cone, consistent with scenarios in which cold clouds survive longer at the interface with the hot wind (e.g., \citealp{villares24}). To illustrate this, in Figure~\ref{massratio} we present a map of the ratio between the mass of the \hi\ to the one of H$\alpha$-emitting gas. We reproject the H$\alpha$ map corrected for extinction to the pixel grid of the \hi\ moment zero map and apply a Gaussian kernel to downgrade it to the resolution of the \hi\ data. We then follow the same procedure described in \cite{watts24} to convert H$\alpha$ surface brightness into mass surface density. In addition to the gas filling factor ($\delta$=0.01), we need to make an assumption on the depth of the outflow cone along the line of sight. Here, we simply use the diameter of a cylinder with volume equal to the volume of the outflow, i.e., $L=$5.5 kpc. This basic assumption would primarily lead to an overestimate of the mass of ionised gas towards the edges of the ionisation cone. Despite this, the map clearly shows that the atomic phase almost always dominates the mass budget (blue), with the H$\alpha$–emitting gas reaching comparable levels only near the centre of the ionisation cone (red).

Of course, Figure~\ref{massratio} should be taken with a grain of salt as the same results would be qualitatively obtained if all the \hi\ was in a thick or radially extended disc and the outflow was only in the warm phase. However, the evidence presented in this paper suggests otherwise. 
Modulo all the assumptions, the \hi\ phase appears to outweigh the ionised one by a factor comparable, or greater than, the difference in their outflow velocities. Consequently, when these quantities are combined to estimate a mass-loading factor, the atomic phase appears to play an important, if not dominant, role. 

Bringing these results together, the properties of the outflow in NGC\,4383 show striking similarities to the well-studied multiphase outflow of M82 \citep[e.g.][]{leroy15, martini18,krieger21,lopez25}. Despite their different optical morphologies, both galaxies share comparable stellar masses and inclinations, and in each case, intense central star formation is likely triggered or enhanced by the gravitational interactions with M81 for M82 \citep{yun94,deblok18}, and within the Virgo cluster environment for NGC\,4383 \citep{chung09}. The similarities extend beyond these global properties: in both systems, the \hi\ component dominates beyond $\sim$1–2 kpc from the disc, with the CO being fully or partially dissociated \citep{krieger21}. The cold gas and dust trace the edges of the ionisation cone, coincident with filamentary H$\alpha$, and PAH emission in the case of M82 \citep{bolatto24,villanueva25}. As in M82, the velocities of the cold component in NGC\,4383 are more consistent with a fountain-like flow than with a large-scale galactic wind.


Our findings add to growing evidence that in the local Universe the cold phase rarely achieves escape velocity or distances larger than a few kpc (e.g., \citealp{marasco23}). Even in extreme cases of bipolar outflows, most of the cold gas appears to remain gravitationally bound and relatively close to the disc, ultimately falling back and re-fueling future star formation. This is in line with the evidence that such outflows typically occur in gas-rich galaxies unlikely to exhaust their reservoirs on short timescales. NGC\,4383 represents a textbook case of this flow, with the role of its multiphase outflow key for the regulating the gas cycling between the disc and the halo. 

Its extended \hi\ disc provides a substantial reservoir into which outflowing gas may eventually fall back, consistent with the galactic fountain scenario \citep{fraternali08}. Thus, galaxies like NGC\,4383 should be viewed as extreme cases of the fountain cycle, in which gas is pushed further from the disc, making the phenomenon easier to trace observationally.

This interpretation also reinforces the likelihood that some of the extraplanar gas we detect is already in the process of returning to the disc. For example, even under a conservative assumption that the neutral gas travels at 100 \kms, it would require only $\sim$50 Myr to reach 6 kpc (the full extent of the ionised outflow corrected for inclination). Unless the starburst is both very young and short-lived, it is therefore likely that at least part of the cold material has already reached its maximum altitude and begun to fall back.

While this complicates attempts to derive reliable mass outflow rates, it also opens an exciting opportunity to directly observe cold material on its way back to the disc, or being recently accreted. Such detections have thus far eluded the community but would provide crucial constraints on the physics of gas cycling in and out of galaxies. Achieving this goal will require tracing the atomic phase at spatial resolutions below $\sim$500 pc, where a full three-dimensional decomposition of the \hi\ structure could be attempted even in galaxies with complex kinematics such as NGC\,4383. Excitingly, the Very Large Array is already making this possible in the Local Group \citep{pingel24}. While we will have to wait for the phase 1 of the Square Kilometre Array to achieve this for large statistical samples at the distance of the Virgo cluster, projects such as MAUVE are ideally positioned to start building a statistical sample of galaxies with well-characterised multiphase outflows. This is an essential step towards clarifying the diversity of conditions that drive fountain flows.

\section{Summary}
\label{sec5}
In this paper, we have presented a multiphase analysis of the star-formation-driven outflow in NGC\,4383, combining ALMA CO(2–1), MeerKAT \hi, and VLT/MUSE data as part of the MAUVE project. Our main findings are as follows:
\begin{itemize}
    \item We find evidence for CO clouds likely entrained in the wind close to the heart of the outflow. However, the CO emission rapidly declines with height, suggesting only a limited contribution of molecular gas to the mass budget of the gas ejected above the disc.
    \item Atomic hydrogen dominates the cold component above $\sim$1 kpc. The \hi\ distribution shows plumes aligned with the ionised gas. However, the extended, warped \hi\ disc complicates unambiguous identification of outflowing gas. Independent dust extinction evidence supports the presence of cold material in the outflow.
    \item All observational evidence supports a scenario in which the cold phase dominates the mass budget of the material ejected a few kpc above the disc.
\end{itemize}

Overall, the multiphase properties of NGC\,4383 reinforce the view that outflows in nearby gas-rich galaxies rarely expel cold gas permanently, but instead cycle material between disc and halo. This highlights the importance of tracing all gas phases at high spatial resolution to fully characterise the baryon cycle in galaxies.

\paragraph{Acknowledgments}
We thank the anonymous referee for helpful comments which improved the quality of this manuscript. We thank Enrico di Teodoro and Tobias Westmeier for useful discussions and Laura Ferrarese for sharing the NGVS images. This work is carried out as part of the MAUVE collaboration (\url{https://mauve.icrar.org/}). 
It is based on observations collected with the MeerKAT telescope under program MKT-22034, at ESO under ESO programms 105.208Y and 110.244E, and makes use of the following ALMA data: ADS/JAO.ALMA\#2023.1.00026.S. ALMA is a partnership of ESO (representing its member states), NSF (USA) and NINS (Japan), together with NRC (Canada), MOST and ASIAA (Taiwan), and KASI (Republic of Korea), in cooperation with the Republic of Chile. The Joint ALMA Observatory is operated by ESO, AUI/NRAO and NAOJ.
The MeerKAT telescope is operated by the South African Radio Astronomy Observatory, which is a facility of the National Research Foundation, an agency of the Department of Science and Innovation. The authors acknowledge the use of services that have been provided by AAO Data Central  (datacentral.org.au) and by the Canadian Advanced Network for Astronomy Research (CANFAR) Science Platform operated by the Canadian Astronomy Data Center (CADC) and the Digital Research Alliance of Canada (DRAC), with support from the National Research Council of Canada (NRC), the Canadian Space Agency (CSA), CANARIE, and the Canadian Foundation for Innovation (CFI). 

This work was supported by resources awarded under Astronomy Australia Ltd's ASTAC merit allocation scheme on the OzSTAR national facility at Swinburne University of Technology. The OzSTAR program receives funding in part from the Astronomy National Collaborative Research Infrastructure Strategy (NCRIS) allocation provided by the Australian Government, and from the Victorian Higher Education State Investment Fund (VHESIF) provided by the Victorian Government.

In addition to the softwares listed in the text, this study has made extensive use of \textsc{Python} \citep{Python3} libraries --- \textsc{Astropy} \citep{Astropy2013, Astropy2018, Astropy2022}, \textsc{\texttt{Numpy}} \citep{Numpy}, \textsc{\texttt{Scipy}} \citep{Scipy}, \textsc{Matplotlib} \citep{Matplotlib},  \textsc{Jupyter} \citep{Jupyter} and \textsc{extinction}.

\paragraph{Funding Statement}
LC and ABW acknowledge support from the Australian Research Council Discovery Project funding scheme (DP210100337). VV acknowledges support from the Comit\'e ESO-CHILE 2024 funding under the name ``Unveiling the physical processes that shape galaxies through cosmic time with ALMA and JWST. TK gratefully acknowledges the financial support from The Leverhulme Trust.
PJ acknowledges support from the project 25-19512L of the Czech Science Foundation and the institutional project RVO:67985815.

\paragraph{Data Availability Statement}
The ALMA, MeerKAT, and MUSE data used in this work are publicly available through the respective observatory archives. ALMA data can be accessed via the ALMA Science Archive (\url{https://almascience.nrao.edu/asax/}) under project ID 2023.1.00026.S; MeerKAT data are available from the South African Radio Astronomy Observatory archive (\url{https://archive.sarao.ac.za/}) under project ID MKT-22034; and MUSE data can be retrieved from the ESO Science Archive Facility (\url{https://archive.eso.org/}) under project ID  105.208Y and 110.244E. The value-added ALMA and MUSE products produced as part of the MAUVE project will be included in forthcoming public data releases and are available from the authors upon request.

\paragraph{Author Contributions}
{\it Conceptualization}: LC, AWB; 
{\it Observations}: JS (ALMA), AWB (MeerKAT), LC, AWB and BC (MUSE); {\it Data Reduction}: JS (ALMA), AWB and SS (MeerKAT), LC and BC (MUSE); {\it Data Products}: LC, TB and JvdS (MUSE), LC (MeerKAT), JS (ALMA); {\it Writing – Original Draft}: LC; {\it Writing – Review \& Editing}: All authors.


\appendix
\renewcommand{\thefigure}{A\arabic{figure}}
\setcounter{figure}{0}
\normalsize

\section{CO(2-1) position-velocity diagram at 500 parsec resolution.}
\begin{figure*}
\centering
\includegraphics[width=.9\linewidth]{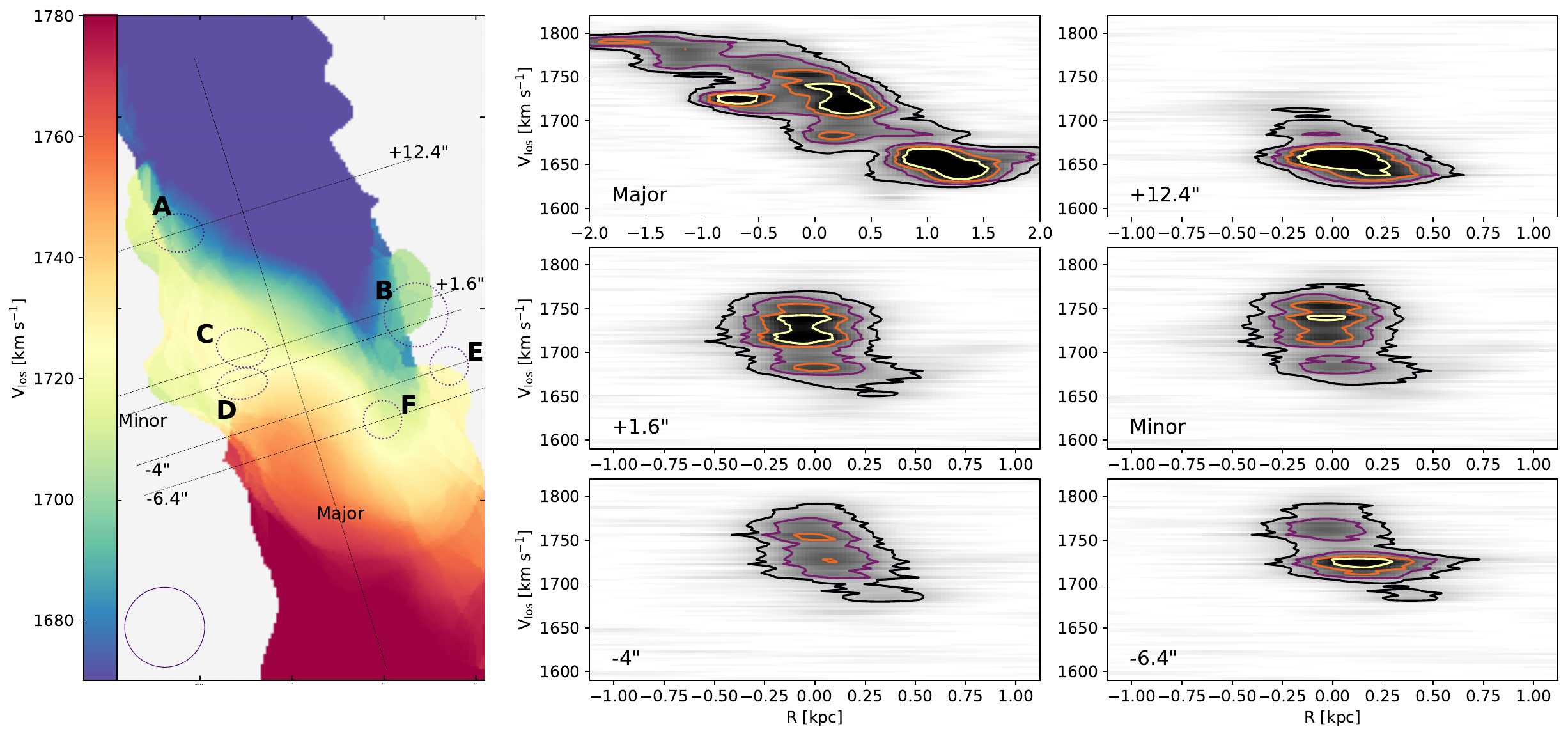}
\caption{{\bf Position-velocity diagrams for the CO(2-1) emission at 500 parsec resolution.} Same as Figure\ref{co_pv} but for a cube degraded to a 500 parsec resolution.}
\label{500pc}
\end{figure*}

To illustrate the dramatic impact of spatial resolution on our ability to identify kinematic disturbances, in Figure~\ref{500pc} we reproduce Figure~\ref{co_pv} for a cube degraded to a resolution of 500 pc. Although this resolution remains higher than that of the MeerKAT data, it is evident that almost all signatures of disturbed kinematics disappear.

\bibliographystyle{apj}
\bibliography{bibliography}

\end{document}